\documentclass[%
 reprint,
 amsmath,amssymb,
 aps,
]{revtex4-2}

\usepackage{graphicx}
\usepackage{dcolumn}
\usepackage{bm}

\usepackage[textsize=tiny,backgroundcolor=red,linecolor=red,textwidth=2cm]{todonotes}

\begin{document}

\preprint{APS/123-QED}

\title[]{Hyperuniform disordered parametric loudspeaker array }
\author{Kun Tang$^{1,2}$, Yuqi Wang$^1$, Shaobo Wang$^1$, Da Gao$^1$, Haojie Li$^3$, Xindong Liang$^3$, Patrick Sebbah$^4$, Yibin Li$^2$,
Jin Zhang$^{1}$}
\email{jinzhang@zhejianglab.com}
\author{Junhui Shi$^{1}$}
\email{junhuishi@zhejianglab.com}

\affiliation{$^1$ Research Center for Humanoid Sensing, Zhejiang Lab, Hangzhou 311100, China.}
\affiliation{$^2$ Tianmushan Laboratory, Xixi Octagon City, Yuhang District, Hangzhou, 310023 China.}
\affiliation{$^3$ Taiji Laboratory for Gravitational Wave Universe, School of Physics and Optoelectronic Engineering, Hangzhou Institute for Advanced Study, Hangzhou, 311100, China}
\affiliation{$^4$ Department of Physics, The Jack and Pearl Resnick Institute for Advanced Technology, Bar-Ilan University, Ramat-Gan 5290002, Israel.}

\date{\today}

\begin{abstract}
A steerable parametric loudspeaker array is known for its directivity and narrow beam width. However, it often suffers from the grating lobes due to periodic array distributions. Here we propose the array configuration of hyperuniform disorder, which is short-range random while correlated at large scales, as a promising alternative distribution of acoustic antennas in phased arrays. Angle-resolved measurements reveal that the proposed array suppresses grating lobes and maintains a minimal radiation region in the vicinity of the main lobe for the primary frequency waves. These distinctive emission features benefit the secondary frequency wave in canceling the grating lobes regardless of the frequencies of the primary waves. Besides, the hyperuniform disordered array is duplicatable, which facilitates extra-large array design without any additional computational efforts.
\end{abstract}

\maketitle

\section{\label{sec:level1}Introduction}

Acoustic beamforming is a signal processing technique used in acoustical arrays for directional sound emission or reception \cite{CHIARI2019}.
Optimal beamforming has been widely researched in the context of sensor arrays \citep{VanVeen1988}. Due to the reciprocity principle of acoustics, these techniques, e.g. design of such microphone arrays and the associated processing algorithms, can also be adopted in the context of loudspeaker arrays \citep{Oliv2016,Ming2015,cheer2012,Mabande} or underwater transducer arrays, bringing applications such as personal audio zone generation \citep{elliott2012,sugi2012}, physical therapies \citep{HIFU2018}. Traditional acoustic arrays have been largely based on various geometrical arrangements (linear, circular, triangular, spiral, and spherical form of arrays) \citep{Mihailo2010,Koya2019,Sladeczek2016,Nordborg,Zebb}, due to mathematical simplicity. Such designs suffer from fundamental restraints, so-called diffraction limits which result in unwanted gratings lobes in both spatial and angular domains, bringing information leakage or harmful physical heating in the off-target regions. To suppress the grating lobes, different optimization techniques have been proposed, e.g. optimizing the array of random disorder \citep{Sarradj,Pernot_2003}, or imposing spatial filters on the input signals \cite{cai2019}. However, the process of optimizing arrays with many elements under certain constraints is a high-dimensional nonlinear problem with many design variables \citep{ARCO2019}. The computational cost would be significant when the number of array sources increases.

The prototype of a parametric loudspeaker was conceived by Westervelt nearly 50 years ago \cite{Westervelt}. It is an application of the parametric acoustic array in air, which generates a low-frequency sound beam from the interaction of two collimated ultrasonic beams as a result of nonlinear acoustic effects \cite{hamilton2008}. The high-frequency ultrasonic components are commonly referred as the primary frequencies and the low-frequency component generated from the interaction of primary frequencies is thus referred as secondary (difference) frequency. Owing to the collimation behavior of ultrasonic beams when propagating in air, the parametric loudspeaker overperforms the traditional loudspeaker in terms of rendering audible sound with high directivity, especially at low frequency range \cite{GAN2012,shi2014}.

To achieve broader applications, such as sound field reinforcement or personal audio space realization \citep{elliott2012,choi2013,sugi2012}, audible sound beam steering using acoustic beamforming techniques \cite{CHIARI2019} is adopted. In a recent work \citep{shi2011}, Shi. \emph{et. al} has successfully developed a steerable parametric loudspeaker array (PLA) using periodic distributions. The results show that the secondary audible wave partially inherits the directive features from the primary waves. Although it is mathematically simple, the periodic configuration suffers from fundamental restraints, so-called diffraction limits which result in unwanted gratings lobes in both spatial and angular domains \cite{Christo2021}. The occurrence of grating lobes for secondary wave found in the periodic distributed PLA \citep{shi2011} generates sound transmitting towards undesired directions and causes disturbances, which therefore should be eliminated by reconfiguring the PLA with new configurations.

The concept of hyperuniform disorder (HUD) was first introduced as an order metric for ranking point patterns according to their local density fluctuations \citep{pre2003}. Hyperuniform structures cover the intermediate regime between random and periodic structures and thus exhibit properties usually associated with both. Hyperuniform stealthy disordered photonic (phononic) structures exhibit large isotropic photonic (phononic) band gaps for the light (elastic wave) for all polarizations \citep{pnas2009,Yu2021,marian2017} and possess rather unusual scattering properties for light \citep{Piechulla2021} and sound waves \cite{elie2022}. Recently, the design of phased antenna arrays with hyperuniform disorder has been proposed in microwave which exhibits directive radiation for large steering angles and wide operating bandwidths \citep{Christo2021}. The radiation patterns of 16 Vivaldi elements have been examined due to the limited number of ports in the measurement system. Due to the different nature of the HUD and the traditional array configurations (linear, circular, triangular, spiral, and spherical form of arrays) \citep{Mihailo2010,Koya2019,Sladeczek2016,Nordborg,Zebb,Sarradj}, we would like to apply this new concept to the design of the PLA to benefit from its distinctive features.

Here we implement the HUD distribution in the design of a PLA \citep{tanaka2010,Takeoka2010} to achieve high-directive rendering of audible sound. Owing to the fact the working frequencies for airborne sound are much lower than microwaves, the bandwidth restraint of the driving electronic systems is no longer a concern. Around 200 elements can be implemented in our experimental setup with low-cost electronic components and transducers \citep{Morales2021,asier2018}. We demonstrate that the proposed array suppresses grating lobes and maintains a minimal radiation region in the vicinity of the main lobe for the primary frequency waves. These distinctive sound emission features benefit the secondary waves in complete cancellation of grating lobes, regardless of the working frequencies of the primary waves \cite{shi2011}. Interestingly, we found that the grating lobes of the secondary waves even at very low frequencies, e.g. a 1kHz audio sinusoidal signal carried by an ultrasonic beam at a frequency 40kHz, can be completely canceled.

\begin{figure*}
\centering
\includegraphics[width = 14cm]{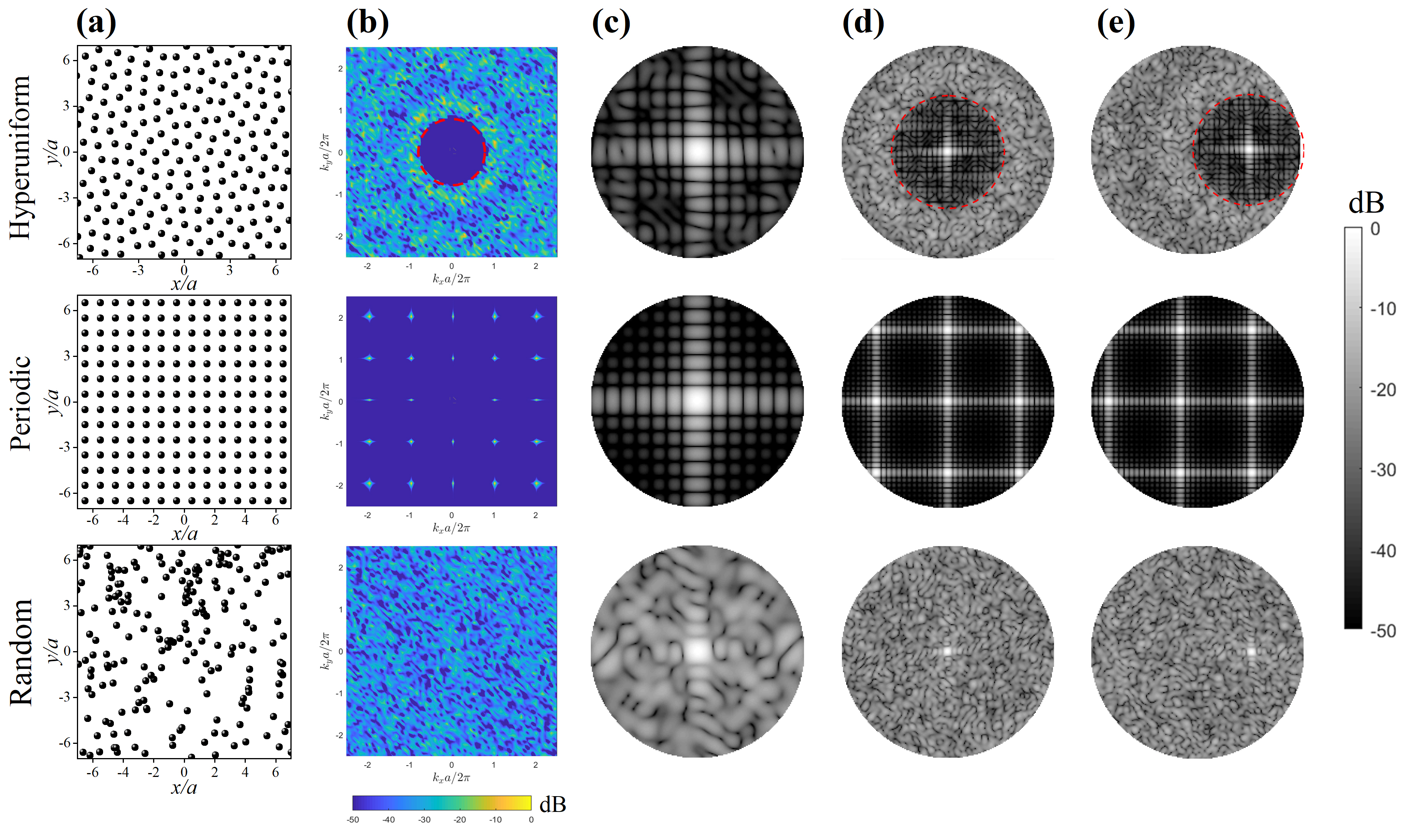}
\caption{\textbf{Structure factor \& Array factor for different point patterns.} (a) Schematics of a HUD pattern in real space with 200 points and $\chi$=0.5 (upper panel), the periodic array (middle panel), and the random array (lower panel). (b) The normalized magnitude of the structure factor in reciprocal space with a logarithmic scale. 
The array factor patterns when $f=f_0$ (c) and when $f=3f_0$ (d) and the main lobe is steered towards $ \phi=0^\circ$, $\theta=30^\circ$ direction (e), measured from the array’s boresight (axis of maximum radiated power), with $f_0$ being the frequency for which the average distance between the nearest elements in the HUD array equals half of the operating wavelength. The angular and radial coordinates in the polar radiation plots correspond to the $\phi$ and $\theta$ angles, respectively.}
\end{figure*}

\section{\label{sec:level1}Radiation properties of HUD acoustic arrays}
We start with a brief explanation of radiation properties for the HUD acoustic array. A hyperuniform point pattern is a point pattern in real space for which the number variance $\sigma^2(R)$ within a spherical sampling window of radius $R$ (in $d$ dimensions) grows more slowly than the window volume ($ \propto R^d $) for large $R$. This concept can be well understood within the reciprocal space, where the structure factor $S(\vec{k})$ for a hyperuniform point pattern can be obtained through a 2D Fourier transform of the point pattern in real space. The resulting structure factor $S(\vec{k})$ vanishes as the wave vector approaches zero $\lvert \vec{k} \rvert \to 0$,
\begin{equation}
\centering
S(\vec{k}) = \frac{1}{N} \lvert \sum_{n=1}^{N} e^{i \vec{k} \cdot \vec{r_n}} \rvert
\end{equation}
where $\vec{k}$ are vectors in reciprocal space and
$\vec{r_n}$ are the positions of the $N$ particles.

We furthermore consider that the structure factor $S(\vec{k})$ is isotropic and vanishes for a finite range of wave numbers $0 < \lvert \vec{k} \rvert \leq k_c$ for the cutoff radius $k_c$ \citep{marian2017}. The size of the vanished region for the structure factor can be expressed by the stealthy parameter
\begin{equation}
\centering
\chi= M(k_c)/Nd,
\end{equation}
where $M(k_c)$ is the number of linearly independent $\vec{k}$ vectors where $S(\vec{k})=0$.

As an example, we illustrate the characteristics of the HUD array with a point pattern consisting of 200 points and $\chi=0.5$, as shown in the upper panel pf Fig.~1(a), $a$ being the average nearest neighbor distance among the points. Fig.~1(b) depicts the structure factor of the hyperuniform point pattern shown in Fig.~1(a). The cut-off radius $k_c$ of the region with vanishing structure factor is estimated with Eq.~(2) and shown in Fig.~1(b) with a red dashed circle. We also compare its structure factor with two typical arrays which possess the same number of points within the occupied area, e.g. the 14$\times$14 periodic array of 196 points and the random arrays of 200 points. It is observed that periodically distributed peaks show up in the structure factor of the periodic array (middle panel) as in the diffraction pattern for a typical crystal, which reveals high symmetry in the point pattern. In contrast, the structure factor of the random array is homogeneously distributed (lower panel). Thus the HUD point pattern with $\chi=0.5$ displays a moderate level of ordered symmetry other than these two extremes.

The normalized far-field radiation pattern in the reciprocal space of an array of identical elements distributed according to a point pattern can be described by its array factor
\begin{equation}
\centering
| A ( \vec{k} )| ^2 = \frac{1}{N} | \sum_{j=1}^{N} e^{i \vec{k} \cdot \vec{r_j}} | ^2,
\end{equation}
where $N$ identical elements are located in the $z$=0 plane of a Cartesian coordinate system at positions $\vec{r_j}$. Here, the wavevector, $k$, is associated with the working wavelength $\lambda$ and the position of the observer in real space, which can be expressed by the corresponding elevation and azimuth angles $(\theta, \phi)$:
\begin{equation}
\centering
\vec{k} = \frac{2 \pi}{\lambda} \sin{\theta} (\cos{\phi}, \sin{\phi}).
\end{equation}

It’s worth noting that despite the similarities between the arithmetic expressions of the structure factor in Eq.~(1) and the array factor Eq.~(3), the physical meanings of the wavevectors are different. The former $\Vec{k}$ is related to the wavevectors in reciprocal space, whereas the latter $\Vec{k}$ (Eq.~(4)) is associated with the spatial position of the observer in real space and working frequencies.

\begin{figure}
\centering
\includegraphics[width = 7cm]{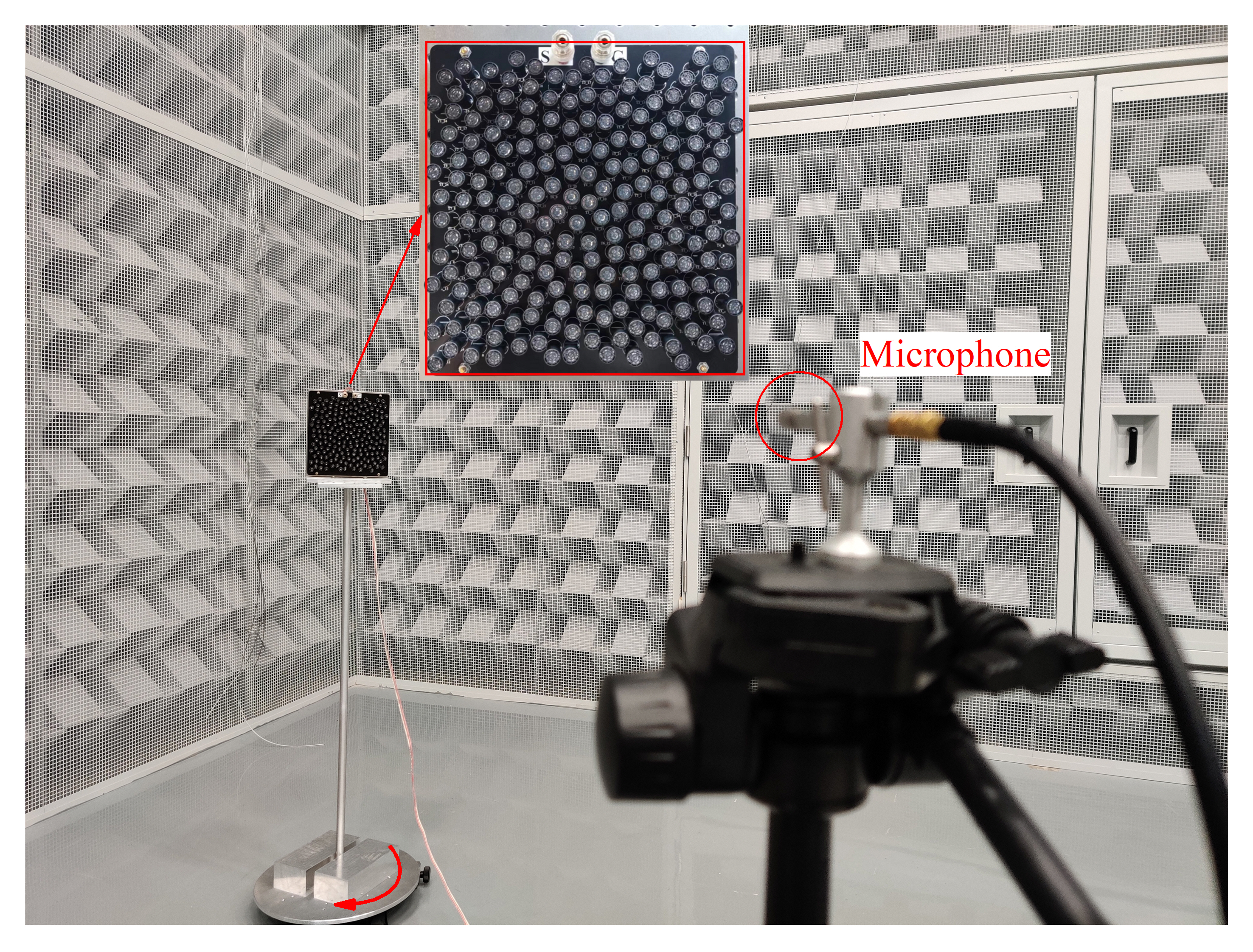}
\caption{\textbf{Experimental setup.} The fabricated 200-element HUD transducer array is mounted on a motorized rotary stage (marked by the red bending arrow) with a standing pole. The sound intensity emitted by the transducer array at each spinning angle is measured by a microphone (marked by a red circle), which is pointed towards the center of the array and fixed at the same height and 2 meters away from the array. Inset: enlarged view of the transducer array. All the measurements are conducted within a semi-anechoic chamber to reduce the unwanted reflections from the surrounding environments.}
\end{figure}

In Fig.~1(c)-1(d), we present the normalized magnitude of the array factors, $A(\vec{k})$, in decibel scale at working frequencies $f_0$ and $3f_0$, where $\lambda_0/2=a$, $\lambda_0$ is the corresponding wavelength of $f_0$. Regarding the far-field radiation for a periodic array, diffraction of the emitted acoustic wave is due to phase differences that result in constructive and destructive interference \cite{Johnson1993}. According to the Nyquist criterion in array signal processing theory, no grating lobes appear at frequency $f_0$ as shown in the middle panel of Fig.~1(c). As soon as the working frequency $f$ is larger than $f_0$, grating lobes appear and can be found in the direction
\begin{equation}
\centering
\theta=\arcsin(\sin{\theta_0} + n*\frac{\lambda}{a}),
\end{equation}
where $\theta_0$ is the steering angle, $n$ is index of grating lobes, and $\lambda$ is the working wavelength at frequency $f$. As shown in the middle panel of Fig.~1(d), grating lobes occur in the direction $\theta=\pm41.8^\circ$ at frequency $3f_0$ and agree with the angle predicted by Eq.~(5). As frequency increases further above $3f_0$, more grating lobes occur and the grating lobes move closer to the main lobe. It's worth noting that no grating lobes show up in the array factor for the HUD array (upper panel of Ffig.~1(e)), which is in stark contrast to the array factor for a normal periodic array. Besides, one can also observe a circular region with very weak radiation that surrounds the main lobe, in contrast to the array factor for a random array (lower panel). This circular exclusion region can act as a protective layer for the main lobe and protects it from exterior interference \cite{Christo2021}. The radius of the circular exclusion region is given by
\begin{equation}
\centering
\theta_{exc}=\arcsin{\frac{k_c \lambda}{2\pi}},
\end{equation}
where $k_c$ is the structure factor cutoff radius and $\lambda$ is the operating wavelength. The estimated cutoff radius $\theta_{exc}=31.8^{\circ}$ obtained from Eq.~(6) for the far-field radiation when $f=3f_0$ is plotted with a red dashed line in the upper panel of Fig.~1(d), which fits the array factor calculated from Eq.~(3). Assuming $\theta_0=0^{\circ}$ and $n=1$ in Eq.~(5), the first-order diffraction for a periodic array would occur in the direction $\theta=\arcsin(\lambda/a)$, which is larger than $\theta_{exc}$ since $k_c<2\pi/a$ (this condition is always ensured in the definition of a HUD array, otherwise the HUD array would collapse into a periodic array). The size of the circular exclusion region for a HUD array is smaller than the angular distance of $\pm1$-order diffraction for a periodic array. As frequency increases, $\theta_{exc}$ decreases as predicted by Eq.~(6), and no grating lobes occur within the exclusion region regardless of the frequency (see the supplementary material of Ref. \cite{Christo2021}). We have presented 5 different HUD point patterns with increasing stealthy parameter $\chi=$0.1-0.5 in Appendix A, accompanied by corresponding simulated structural factors and array factors. As $\chi$ increases from 0.1 to 0.5 gradually,  $M (k_c)$ grows as well, which ends up with the expansion of the circular exclusion region as predicted by Eq.~(6). Generally, the point pattern is considered to be in the disordered regime when $\chi \leq 0.5$, but this threshold value may vary with the number of points $N$ \citep{Uche2004}. The HUD point pattern with $\chi=0.5$ still resides in the disordered regime and maintains a relatively broad circular exclusion region at the same time.

Moreover, this unique radiation property of the HUD acoustic array is preserved during beam steering. Fig.~1(e) exhibits the array factor at frequency $f=3f_0$ when the main lobe is steered towards $\phi=0^{\circ}$, $\theta=30^\circ$ direction. The radiation exclusion region surrounding the main lobe of the HUD array is preserved and it also acts as an indication of the steering direction (upper panel). This contrasts with the steered radiation pattern of the random array (lower panel), where the direction of the main lobe is barely distinguishable from the surroundings. As for the case of the periodic array (middle panel), the presence of strong grating lobes makes it difficult to determine the steering direction, which is the so-called spatial aliasing effect \cite{spatial2006,shi2011}. Therefore, we expect the HUD array benefits from the advantages of both periodic (exclusion regions) and random arrays (cancellation of grating lobes) and incorporate them into a single design, thus performing better than both.

\section{\label{sec:level1} Measured directivies of PLA with Hyperuniform disorder}
To further validate the theoretical analysis, we conduct a series of measurements using ultrasound transducer arrays. The typical working frequency is in the range of 20-100kHz and does not require wide bandwidths on transmission electronics, such that we can employ a much larger number of array elements than that operating at microwave frequencies \cite{Christo2021}. Specifically, we fabricate two configurations: a periodic array with 196 elements and a HUD array with 200 elements and $\chi=0.5$, using 10-mm-diameter piezoelectrically actuated transducers (MA40S4S, MURATA, Japan) as the array element for both configurations. The 200-element HUD transducer array along with the test environment is shown in Fig.~2. The overall size of the transducer array is around 200$\times$200 $mm^2$. The far-field radiation directivity of the transducer array is obtained by rotating the planar array and measuring the signal using a microphone placed 2$m$ away from the array center.

\begin{figure}
\centering
\includegraphics[width = 8cm]{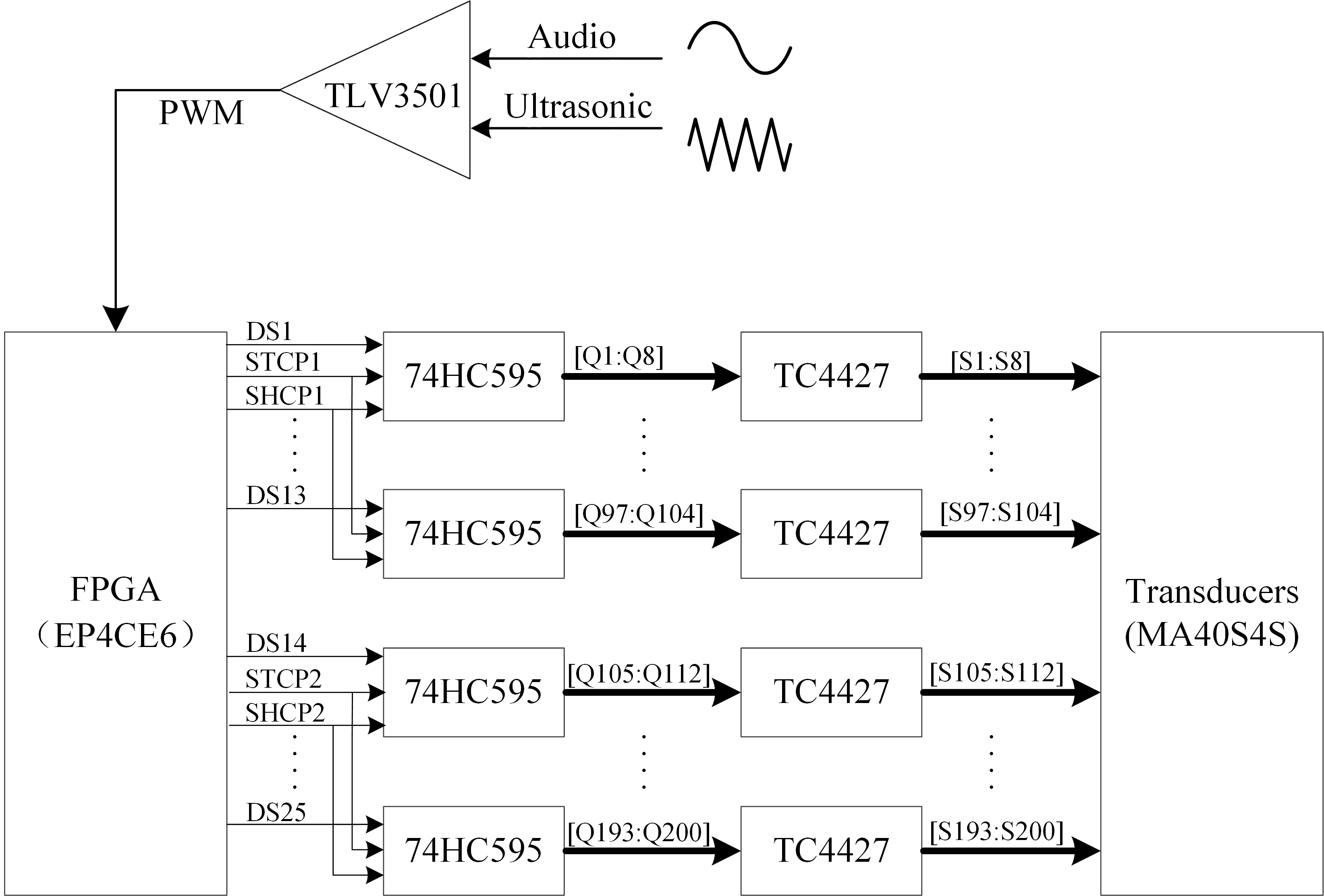}
\caption{\textbf{Block diagram of the PLA.}}
\end{figure}

A PLA is known for its sharp directivity due to the ultrasonic carrier wave. An audio signal modulated onto this carrier wave is reproduced along the beam by intrinsic non-linearity of the air \citep{Takeoka2010}. In order to steer the main lobe towards a specific direction, namely the elevation angle $\theta_s$ and the azimuth angle $\phi_s$, we need to multiply the signal from each element in the array with a complex phase $w_j=e^{-i \vec{k_s} \cdot \vec{r_j} }$, where $\vec{k_s}$ is the steered wavevector defined as $\vec{k_s} = \frac{2 \pi}{\lambda} \sin{\theta} (\cos{\phi}, \sin{\phi})$, $\vec{r_j}$ is the position vector of the $j$th array element, $\lambda$ being the working wavelength. In the experimental setup, the phase differences between each element are controlled by signal delays. As shown in Fig.~3, the desired audio signal is first compared with a 40kHz triangle wave to generate a pulse-width modulated (PWM) signal by a comparator chip (TLV3501AIDR, TI). Then the PWM signal is input to the Field Programmable Gate Array (FPGA, EP4CE6E22C8N, Intel), which performs digital delay for the 200 channels of external PWM signal. The resolution for the signal delay of each channel is 0.8 $\mu{s}$, which guarantees the accuracy of beam steering \cite{WU2012}. Each delayed PWM signal is amplified from 5V up to 12V using a dual MOSFET (TC4427AEOA713, Microchip) driver. A serial-to-parallel chip (74HC595D,118, Nexperia) is employed to reduce the number of output pins. Through
the above analog modulation, the amplitude-modulated ultrasound wave has a carrier, upper and lower side-band components, which generate the audible sound in the air due to the nonlinear interaction of the carrier component and each side-band component in the ultrasound beams.

\begin{figure}
\centering
\includegraphics[width = 8.5cm]{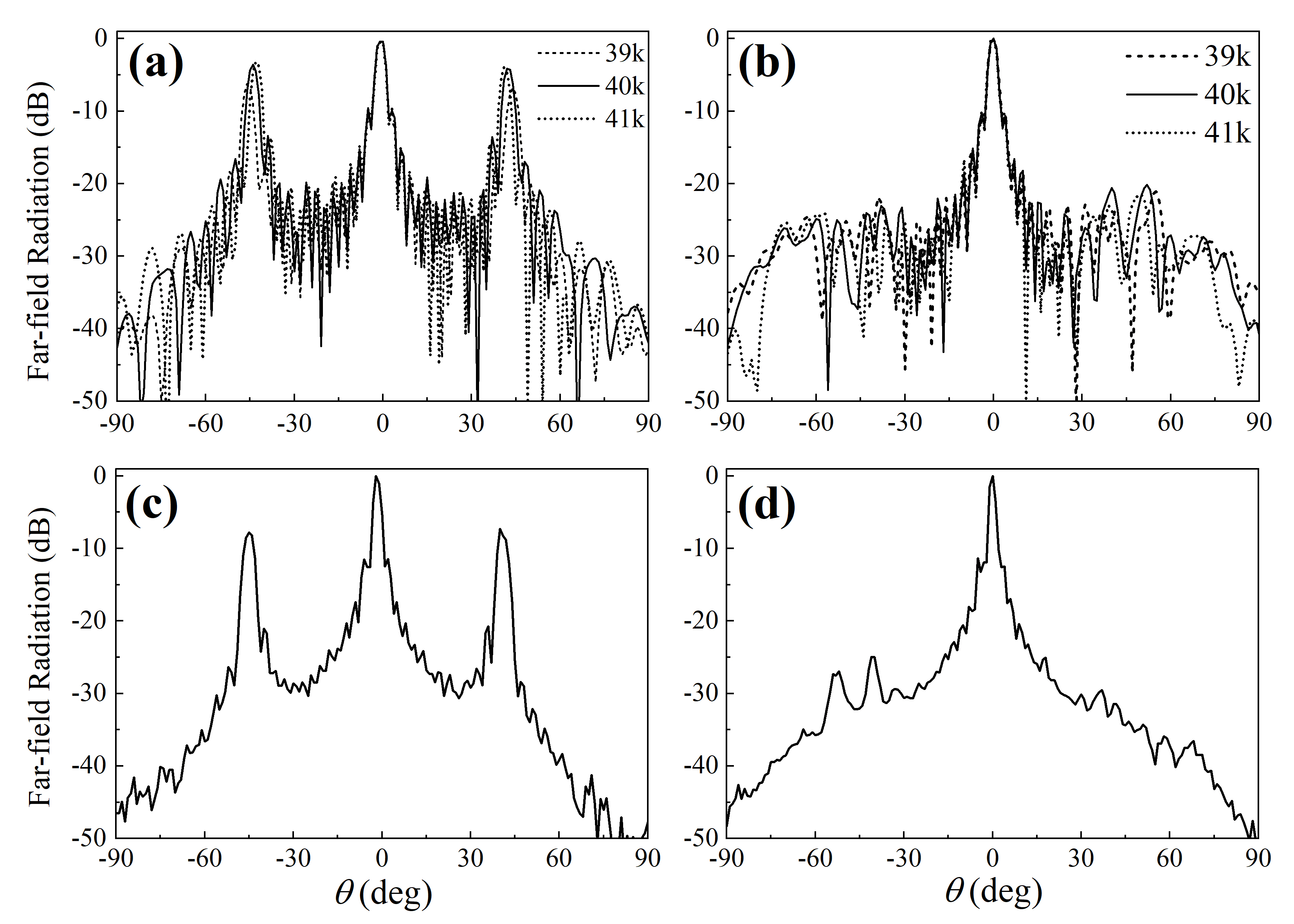}
\caption{\textbf{Forward far-field directivity patterns of PLA.} Measured far-field directivity in the azimuth plane ($\phi=0^{\circ}$) for primary frequencies 39kHz (black dashed line), 40kHz (black solid line), and 41kHz (black dotted line) for the periodic array (a) and the HUD array (b). The corresponding measured directivity of the secondary wave at 1kHz is plotted in (c) and (d), respectively.}
\end{figure}

Note that the proposed HUD array is short-range random while correlated at large scales. The inter-emitter distance in a HUD array fluctuates locally. To avoid spatial overlap of the transducers, we set the average nearest neighbor distance (center to center) to be 12.75mm, which is 3/2 of the wavelength at the transducer resonance frequency of 40kHz and larger than the diameter of the emitter (10mm). To carry out a fair comparison with the HUD array, we set the period of the periodic array equal to 12.75mm as well. We conduct our experiments near the resonance frequency of the transducers $f$ = 40kHz, which is equal to $3f_0$. A sinusoidal wave signal at frequency 1kHz is modulated to the carrier frequency of 40kHz. The measured far-field directivity patterns in the azimuth plane ($\phi=0^{\circ}$) for primary frequency waves at three different frequencies, 39kHz (black dashed line), 40kHz (black solid line), and 41kHz (black dotted line) for the periodic array are illustrated in Fig.~4(a). We can clearly observe that the main lobe of three different primary frequency waves coincide in the same direction  ($\theta=0^\circ$) with each other, whereas the grating lobes appear in slightly shifted different directions (near $\theta=\arcsin{2/3}=\pm41.8^\circ$) because of the difference between their wavenumbers. The corresponding results for the HUD array are given in Fig.~4(b). As can be seen, the grating lobes of the three primary frequency waves are completely suppressed, and an exclusion region with minimal radiation as low as -30dB ($\lvert\theta\rvert\leq31.8^{\circ}$) can be observed surrounding the main lobe. In Fig.~4(c), we present the measured far-field directivity patterns of the secondary frequency wave at $f$=1kHz (blue solid line) for the periodic array. It’s observed that, besides the main lobe, the grating lobes occur at the direction where the grating lobes of the primary frequency waves overlap (Fig.~4(a)). It comes from the nonlinear interactions of the grating lobes of the carrier component ($f=$40kHz) and each side-band component ($f=$39kHz, 41kHz). The corresponding results of the secondary frequency wave for the HUD array are given in Fig.~4(d). It’s worth noting that the grating lobes are completely canceled, even for such a low-frequency (1kHz) secondary frequency wave carried by an ultrasonic beam (40kHz), which is in stark contrast with the periodic configurations (Fig.~4(c)).

\begin{figure}
\centering
\includegraphics[width = 8.5cm]{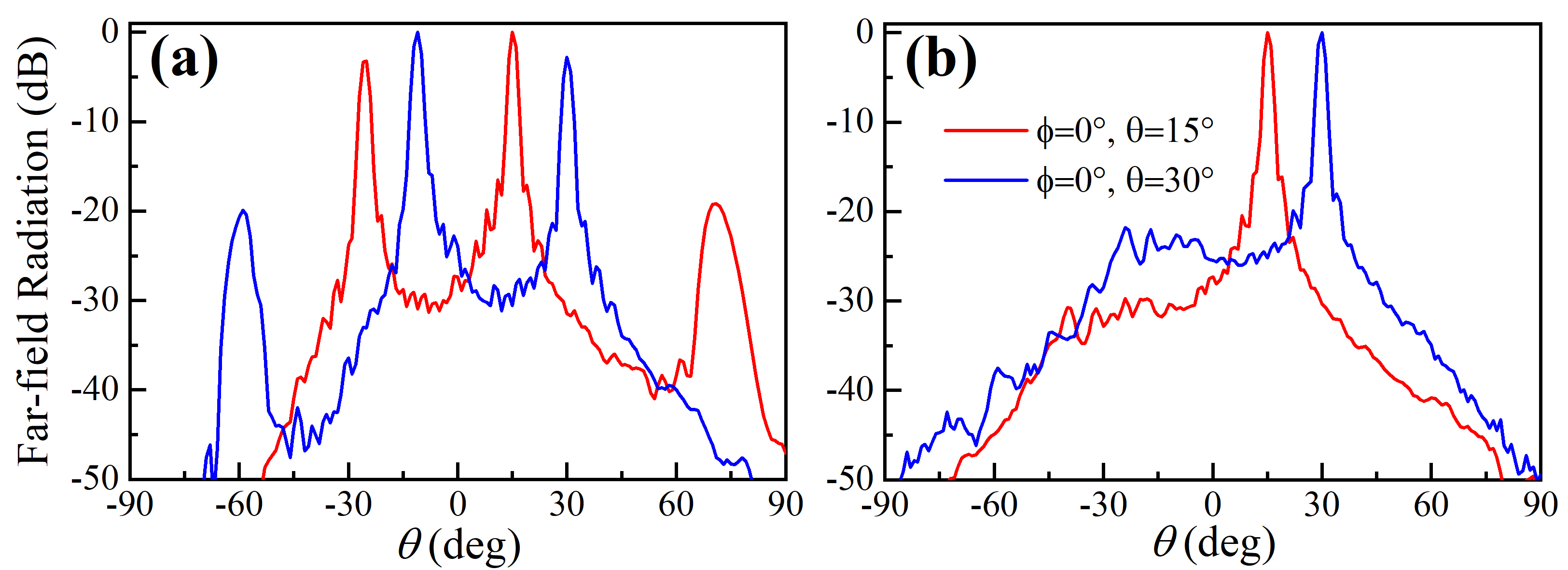}
\caption{\textbf{Steered Far-field directivity patterns of PLA.} Measured directivity of the secondary frequency waves at 1kHz in the azimuth plane ($\phi=0^\circ$) when the main lobe is steered towards the $\phi=0^\circ$, $\theta=15^\circ$ (red solid line), and the $\phi=0^\circ$, $\theta=30^\circ$ (blue solid line) directions for the periodic array (a) and the HUD array (b).}
\end{figure}

We also steered the main lobe of the secondary wave at frequency 1kHz towards two different directions in the azimuth plane, namely the $\phi=0^\circ$, $\theta=15^\circ$ and $\phi=0^\circ$, $\theta=30^\circ$ directions. The measurement results in the azimuth plane ($\phi=0^\circ$) are illustrated in Fig.~5(a) for the periodic array and in Fig.~5(b) for the HUD array. The far-field directivity of the secondary frequency wave for the HUD array is delineated by a main steered lobe, whereas the grating lobes are completely canceled. This contrasts with the radiations for the periodic array where the spatial aliasing effect takes place, e.g. steering direction of the main lobe is hardly distinguished from the grating lobes. We also tested the case of a period of 10mm as in reference \citep{shi2011}, where emitters are touching each other. Although sidelobes reduce with the reduced period, our proposed HUD array still outperforms the periodic array, as shown in Fig. 10 of Appendix C. To summarize the above analysis, the grating lobes of the secondary frequency wave of PLA can be completely eliminated by a HUD array configuration, regardless of frequencies of the primary waves \citep{shi2011}. Our proposed HUD array outperforms the periodic array \textbf{with a period of 12.75mm as well as 10mm} in suppressing the side lobes. Moreover, the exclusion regions around the main lobe are well preserved for the primary frequency waves of the HUD array distributed PLA, although, which is not evident for the secondary frequency wave.

Furthermore, extra-large acoustic arrays with a great number of array elements are required in demanding application scenarios, like high-intensity sound radiations \cite{Kacz2015} or precise source localization \cite{large_array}. Optimization methods proposed to improve the array performances \cite{ARCO2019,Zhu2021} are, however, computationally expensive during the adaption of an extra-large array. Therefore, it is important yet necessary to demonstrate another feature of the HUD array, its duplicability. This behavior inherits from the generation process of a HUD point pattern where periodic boundary conditions are applied to the two-dimensional computational domain \cite{Leseur16}. Thus a large array consisting of periodic replication of a HUD subarray remains hyperuniform disordered and possesses the same distinctive emission behaviors \cite{Christo2021}. We compare the far-field directivity for primary frequency wave at $f=$40kHz of a large array (800 elements) made of a 2$\times$2 subarray with the measured far-field directivity of the subarray (200 elements). Throughout the measurements, we found that the radiation properties of the transducers can be well captured by the piston source model \citep{asier2018} (see the agreement between the simulated and measured directivity for primary frequencies at $f=$30\&40kHz in Fig.~8\&9 of Appendix B). We multiply the radiation of an individual single element with the array factor of the large array, and the resulting directivity patterns are illustrated in Fig.~6. We present the case where no steering is applied in Fig.~6(a) and the case of beam steering towards the $\phi= 0^{\circ}$, $\theta= 30^{\circ}$ direction in Fig.~6(b). As can be seen, for both cases, the large array of 800 elements behaves similarly to the HUD subarray of 200 elements, as the main lobe is surrounded by the exclusion region with the absence of the grating lobes. Besides, the nulling in the exclusion region of the sound radiation for the large array is much deeper than that for the original subarray. With infinite replications of the original subarray, the radiation values in the exclusion region would approach zero as predicted by the structure factor shown in Fig.~1. It is expected that the large array will suppress the grating lobes of the secondary frequency wave as the original HUD subarray considering the above radiation properties for the primary frequency wave.

\begin{figure}
\centering
\includegraphics[width = 8.5cm]{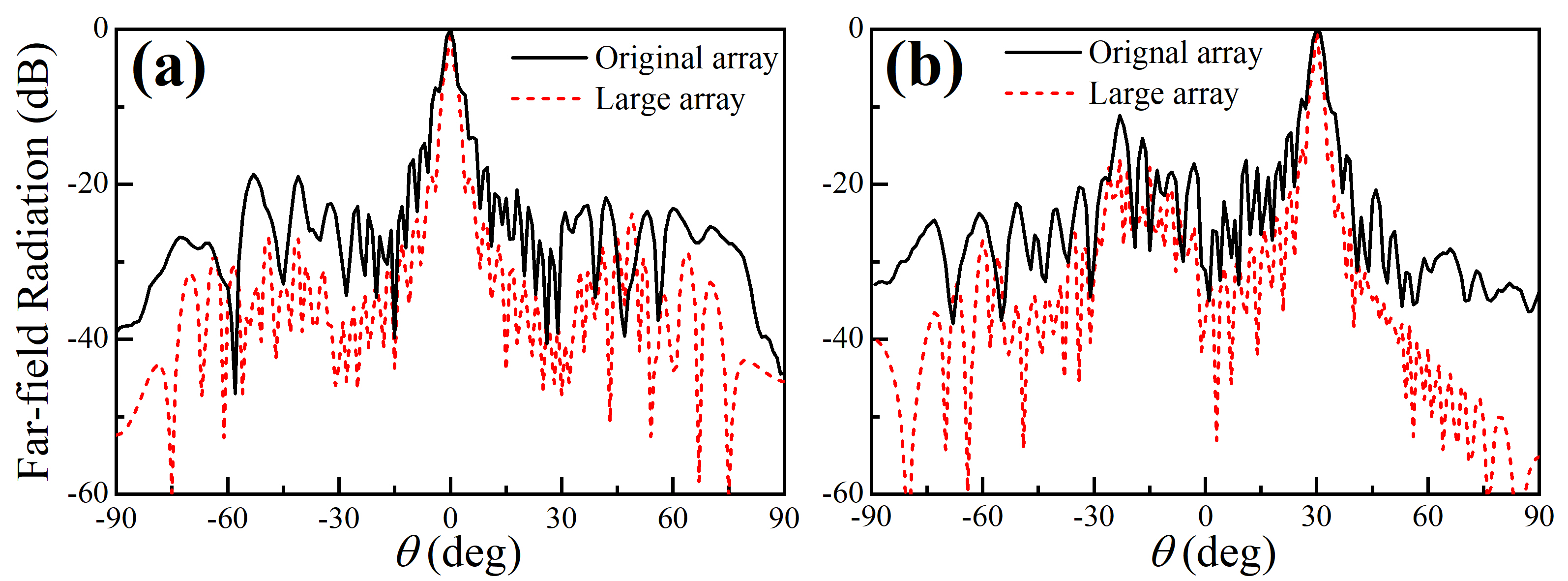}
\caption{\textbf{Far-field directivity patterns for large HUD array.} Simulated directivity for the large HUD array (red dashed line) and measured directivity for the original HUD sub-array (black line) in the azimuth plane ($\phi= 0^{\circ}$) (a), when the main lobe is steered towards the $\phi= 0^{\circ}$, $\theta= 30^{\circ}$ direction (b), at frequency $f=$40kHz.}
\end{figure}

\section{\label{sec:level1}Conclusions}
In conclusion, we developed a PLA which follows a HUD array configuration. Unlike existing acoustic arrays, this distribution originates from an order metric (hyperuniformity) to characterize local density fluctuations of a point pattern. Both the simulated and measured results reveal the effectiveness of the proposed array for the primary frequency waves in suppressing the grating lobes (like a random array) while maintaining a minimal radiation region around the main lobe (like a periodic array), which incorporates them into a single design and performs better than both its periodic and random counterparts. These properties benefit the secondary frequency waves in canceling the grating lobes regardless of the frequencies of primary waves unlike in \citep{shi2011}. Moreover, these HUD arrays are duplicatable to generate extra-large arrays, which avoids exponentially increasing computational costs commonly found by adopting optimizing algorithms \citep{haupt1994,bray2002}. The proposed approach can also be employed for ultrasonic transducer arrays \cite{cai2019} working in other media, e.g. water and human tissues, which could bring potential applications in undersea communication and medical therapies. This design also opens an interesting route to bionic acoustic arrays inspired by the hidden symmetry from nature \citep{avian2014}.

\section*{\label{sec:level1}Appendix A: Radiation property of HUD array as $\chi$ vary}

\begin{figure*}
\centering
\includegraphics[width = 14cm]{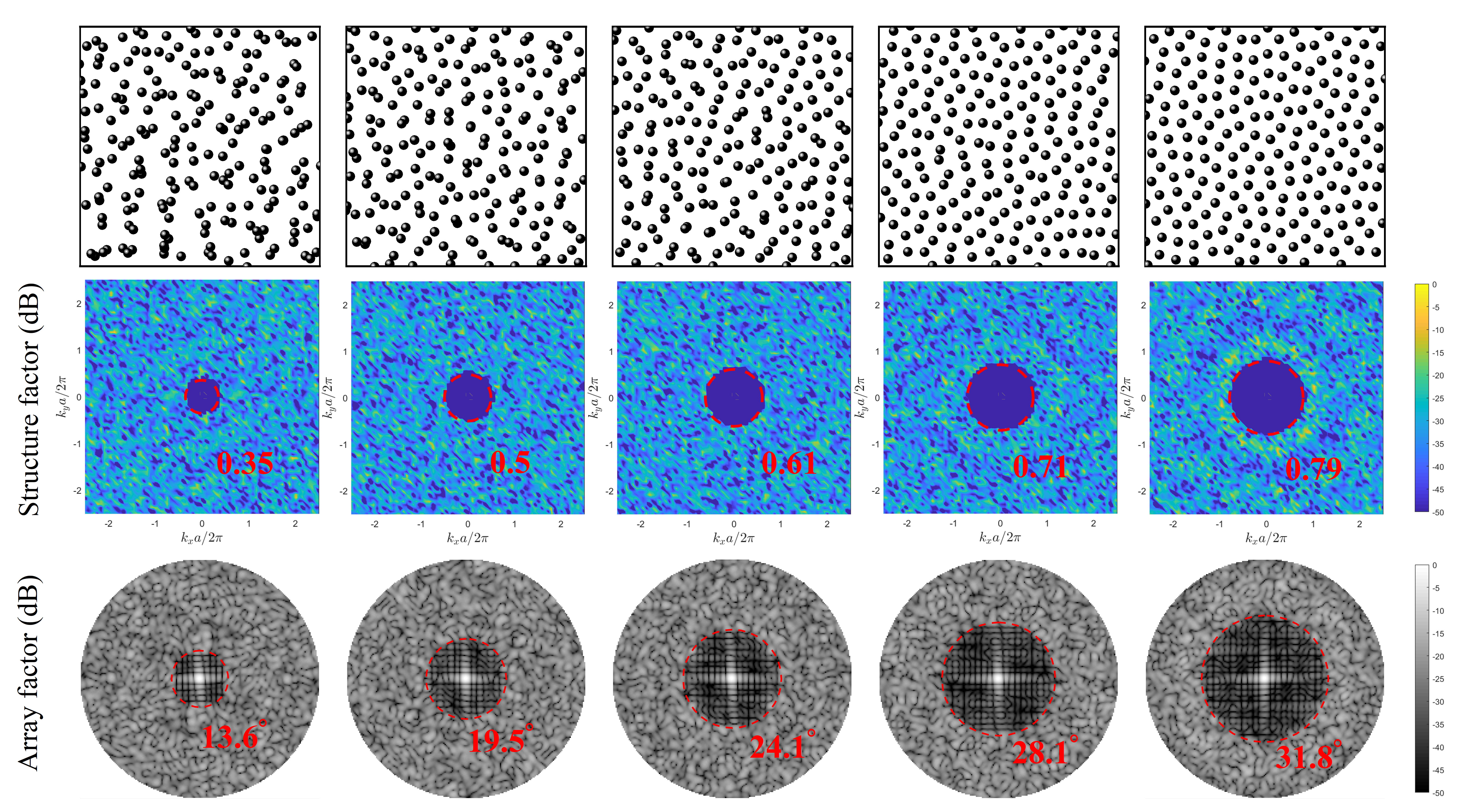}
\caption{\textbf{Radiation property of HUD array as $\chi$ vary.} Schematics of HUD point patterns with different stealthy parameters $\chi$=0.1-0.5 (upper panels); Corresponding structure factor (middle panels), where the red dashed circle indicates the cutoff radius $k_c$ (normalized by $2\pi/a$), the structure factor $S(\vec{k})$ vanishes for $0<\lvert \vec{k} \rvert \leq k_c$; Corresponding array factor (lower panels), where the red dashed circle indicates the radius of the circular exclusion region $\theta_{exc}$.}
\end{figure*}

In the experimental investigation and numerical analysis presented in the main text, the stealthy parameter of the HUD acoustic array was chosen to be $\chi$=0.5. Here, we check the dependence on $\chi$ of the radiation property. As illustrated in the upper panels of Fig.~7, we present a point pattern with the same number of points $N$=200, but with different stealthy parameters $\chi$, ranging from 0.1 to 0.5. One can observe that as $\chi$ grows, the particle clustering effect disappears gradually and occupy the entire space uniformly. The structure factor $S(\vec{k})$ is calculated through a 2D Fourier transform of the point pattern and shown in the middle panels. As $\chi$ grows, the nulling region surrounding the origin $k$=0 where $S(\vec{k})$=0 enlarges and shows great agreement with the calculated $k_c$ (red dashed circle). The corresponding normalized magnitude of the array factors in decibels is plotted in the lower panels, with 0 dB meaning the maximum value. The working frequency is chosen to be 3$f_0$. Similar to the nulling region in the structure factor, the size of the circular exclusion region enlarges as $\chi$ grows and shows great correspondence with the estimated $\theta_{exc}$ (red dashed circle) given by Eq.~(5). It indicates that the point pattern with $\chi$=0.5 possesses the largest exclusion region and suppresses the grating lobes at the same time. Generally speaking, the point pattern is considered to be in the disordered regime when $\chi\leq$0.5. As $\chi$ grows further, the point pattern turns into ordered and the grating lobe takes place like in the periodic array.

\section*{\label{sec:level1}Appendix B: Measured and simulated directivity for primary frequency waves}

\begin{figure}
\centering
\includegraphics[width = 8.5cm]{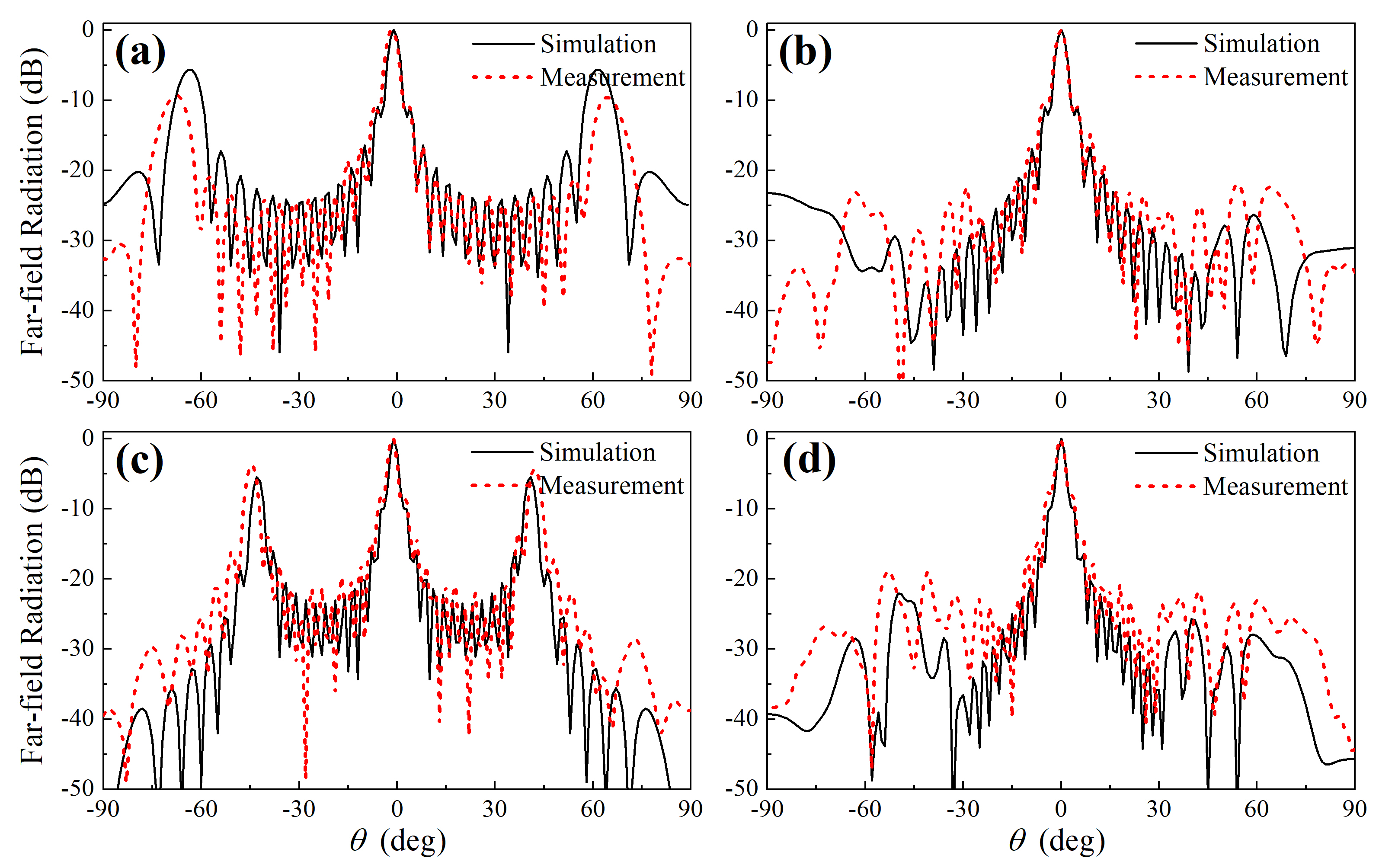}
\caption{\textbf{Forward far-field directivity patterns.} Simulated (black solid line) and measured (red dotted line) far-field directivity patterns in the azimuth plane ($\phi=0^\circ$) when $f$=30kHz (upper panel) and $f$=40kHz (lower panel) for the periodic transducer array (a, c) and the HUD transducer array (b, d).}
\end{figure}

The simulated far-field directivity for the primary frequency wave of
different acoustic arrays are obtained using an analytical model, in which the total radiation pattern of an array with identical source elements can be expressed by the multiplication of the individual single source element with the array factor.
The diameter of the ultrasonic transducer is around 10mm and larger than the working wavelength of 8.5mm at frequency 40kHz. Therefore each emitter cannot be assumed as omnidirectional and can be well modeled as a piston source \cite{asier2018}, which is Fraunhofer diffraction approximation (Fourier transform of a disc aperture or a piston). The complex acoustic pressure at point $\vec{r}$ due to a piston source emitting at a single frequency can be modeled as
\begin{equation}
\centering
P(\vec{r})=P_0 A \frac{D_f(\theta)}{d} e^{i(\phi+kd)},
\end{equation}
where $P_0$ is a constant that defines the transducer amplitude power and $A$ is the peak-to-peak amplitude of the excitation signal. $D_f (\theta)$ is a far-field directivity function that depends on the angle $\theta$ between the transducer normal and $\vec{r}$. Here, $D_f (\theta)=2J_1 (ka \sin\theta)/ka \sin \theta$, which is the directivity function of a circular piston source, where $J_1$ is a first-order Bessel function of the first kind and $a$ is the piston radius. The term $1/d$ accounts for divergence, where $d$ is the propagation distance in free space. $k = 2 \pi/\lambda$ is the wavenumber and $\lambda$ is the wavelength. $\phi$ is the initial phase of the piston.

We conduct our experiments at two different working frequencies $f$ = 30kHz and 40kHz, which are larger than $2f_0$ (26.7kHz) and exactly $3f_0$. The measured and simulated far-field directivity patterns in the azimuth plane ($\phi=0^\circ$) for the periodic array are illustrated in Fig.~8(a) ($f$=30kHz) and Fig.~8(c) ($f$=40kHz). The corresponding results for the HUD array at the same working frequencies are given in Fig.~8(b) and Fig.~8(d). In both cases for different frequencies, the measurement and simulation results agree well with each other within the main lobe region. The discrepancies between the measurement and simulation data outside the main lobe region might be attributed to the misalignment of the motorized stage and the transducer array in the experimental setup. We can observe that, for both working frequencies, the HUD transducer array suppresses the grating lobes and has significantly reduced peak side lobe level values. Specifically, the measured peak side lobe level (PSLL) value at frequency 30kHz (40kHz) is reduced from -9.3dB (-3.8dB) for the periodic array to -21.8dB (-18.7dB) for the HUD array.

\begin{figure}
\centering
\includegraphics[width = 8.5cm]{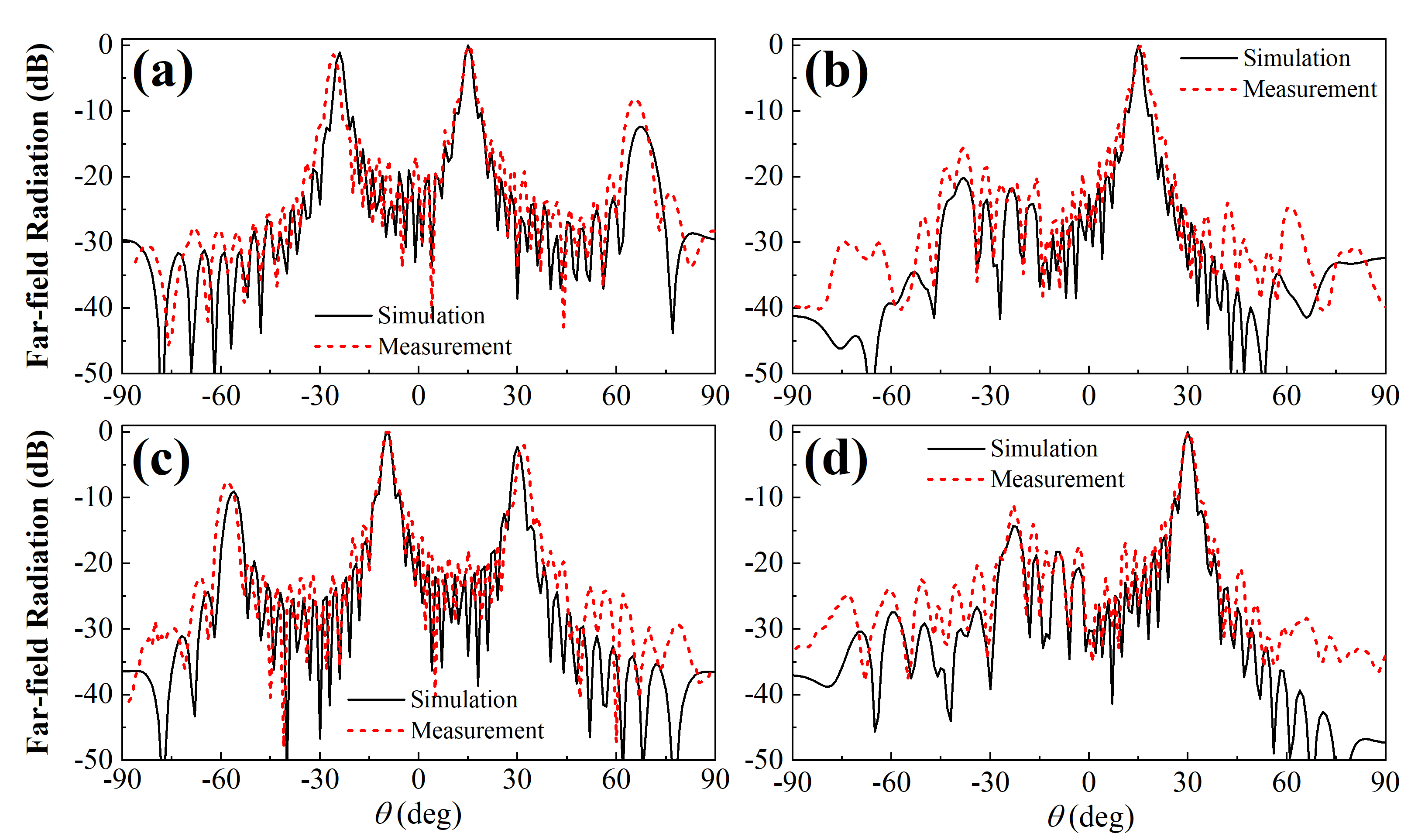}
\caption{\textbf{Steered far-field directivity patterns.} Simulated (black solid line) and measured (red dotted line) far-field radiation patterns in the azimuth plane ($\phi=0^{\circ}$) when the main lobe is steered towards the $\phi= 0^{\circ}$, $\theta= 15^{\circ}$ (a, b) and the $\phi= 0^{\circ}$, $\theta= 30^{\circ}$ (c, d) directions when $f$ =40kHz  for the periodic transducer array (a, c) and the HUD transducer array (b, d).}
\end{figure}

At $f=$40kHz, we also steered the main lobe towards two different directions in the azimuth plane, namely the $\phi= 0^{\circ}$, $\theta= 15^{\circ}$ and $\phi= 0^{\circ}$, $\theta= 30^{\circ}$ directions. The measurement (red dashed lines) and simulation results (black solid lines) in the azimuth plane ($\phi= 0^{\circ}$) are illustrated in Fig.~9(a)\&9(c) for the periodic array and in Fig.~9(b)\&9(d) for the HUD array, which shows great coincidence with each other. As predicted by the simulations for the array factor with steered directions, the far-field directivity pattern for the HUD array is delineated by the main beam surrounded by a weak emission exclusion region, whereas outside this exclusion region the sidelobes are kept at a low level. This is in stark contrast to the far-field directivity pattern for the periodic array where several grating lobes can be seen and steering direction is hardly determined. The maximum side lobe level is even higher than the main lobe when steered towards the $\phi=0^{\circ}$, $\theta=30^{\circ}$ direction. It's worth noting that the measured PSLL value, when steered towards the $\phi=0^{\circ}$, $\theta=15^{\circ}$ ($\phi=0^{\circ}$, $\theta=30^{\circ}$) direction at frequency 40kHz is reduced from -1.4dB (2dB) for the periodic array to -15.4dB (-11.1dB) for the HUD array.

\section*{\label{sec:level1}Appendix C: Measured directivity for a periodic array with a period of 10mm}

We measured the far-field radiation patterns of the
secondary frequency wave at 1kHz for a periodic array \textbf{with a period of 10mm}. As shown in Fig.~10(a), the side lobes of the secondary frequency wave in the forward directivity (black line in Fig.~10(a)) are smaller with a reduced inter-emitter distance, whereas they are still higher than the side lobes for a HUD array (Fig.~4(d) of the main text). Besides, when the main lobe is steered towards the $\phi=0^\circ$, $\theta=15^\circ$ (red line in Fig.~10(b)), the side lobes of the secondary wave are within -10dB less than the main lobe. The level increase of the grating lobe became significant, and even surpass the level of the main lobe  when steered towards the $\phi=0^\circ$, $\theta=30^\circ$ direction. This is in stark contrast with the HUD array where the sidelobes remain at a low level (less than -20dB in Fig.~5 of the main text).

\begin{figure}
\centering
\includegraphics[width = 8.5cm]{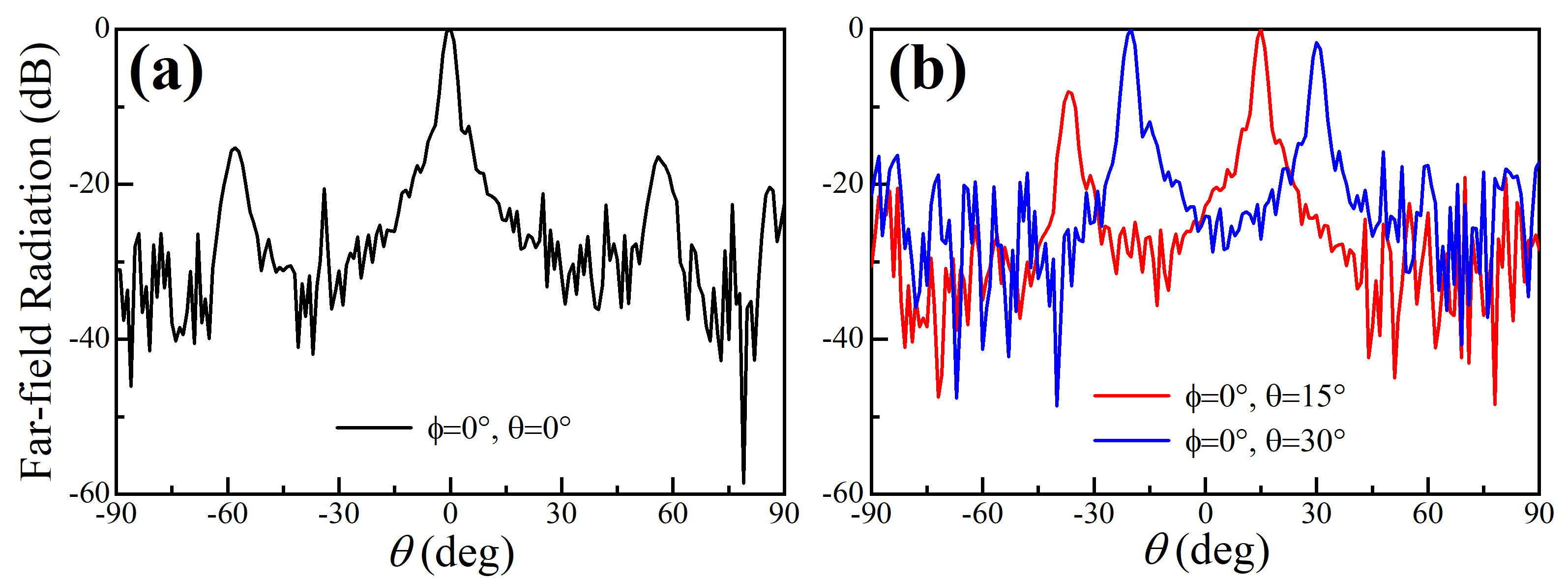}
\caption{\textbf{Far-field directivity patterns of a periodic PLA with a period of 10mm.} Measured far-field directivities of the secondary frequency waves at 1kHz in the azimuth plane ($\phi=0^{\circ}$), for the forward radiation (black line in (a)), when the main lobe is steered towards the $\phi=0^\circ$, $\theta=15^\circ$ (red line in (b)), and the $\phi=0^\circ$, $\theta=30^\circ$ (blue line in (b)).}
\end{figure}

\begin{acknowledgments}
The authors thank Prof. Marian Florescu for generating the initial hyperuniform disordered point patterns. This research was supported by the Youth Foundation Project of Zhejiang Lab (Grant No. 2020MC0AA07). P. S. is thankful to the Israel Science Foundation (Grants No. 1871/15, 2074/15, and 2630/20), and the United States-Israel Binational Science Foundation NSF/BSF (Grant No. 2015694 and No. 2021811).
\end{acknowledgments}

\bibliography{cas-refs}

\end{document}